\documentclass[journal]{IEEEtran}
\usepackage{mathtools}

\usepackage{fancyhdr}
\usepackage{amsmath,amssymb}
\usepackage{amsfonts}
\usepackage[dvips]{graphicx}
\usepackage{dsfont}
\usepackage{comment}
\usepackage{algorithmic}
\usepackage{comment}
\usepackage{caption}
\usepackage{subcaption}
\usepackage{balance}
\usepackage{algorithm}
\usepackage{float}
\usepackage[hidelinks]{hyperref}    

\usepackage{color}
\usepackage{pgfplots}           
	\pgfplotsset{compat=1.16}

\usepackage{enumerate}
\usepackage{graphics}
\usepackage{cite}
\usepackage{mathrsfs}
\usepackage{stfloats}

\usepackage{tikz}
\usepackage{mathdots}
\usepackage{yhmath}
\usepackage{cancel}
\usepackage{color}
\usepackage{siunitx}
\usepackage{array}
\usepackage{multirow}
\usepackage{textcomp}
\usepackage{gensymb}
\usepackage{tabularx}
\usepackage{booktabs}
\usetikzlibrary{fadings}
\usetikzlibrary{patterns}
\usetikzlibrary{shadows.blur}
\usepackage{graphicx}

\newcommand{\pn}{\sigma_{\phi}}

\newtheorem{theorem}{Theorem}

\newtheorem{remark}{Remark}

\begin{document}

\title{Constellation Design and Detection under Generalized Hardware Impairments}
\author{Thrassos K. Oikonomou,~\IEEEmembership{Graduate Student Member,~IEEE}, Dimitrios Tyrovolas,~\IEEEmembership{Member,~IEEE}, \\ Sotiris A. Tegos,~\IEEEmembership{Senior Member,~IEEE}, Panagiotis D. Diamantoulakis,~\IEEEmembership{Senior Member,~IEEE}, \\ Panagiotis Sarigiannidis,~\IEEEmembership{Member,~IEEE}, and George K. Karagiannidis,~\IEEEmembership{Fellow,~IEEE}
\thanks{T. K. Oikonomou, D. Tyrovolas, S. A. Tegos, P. D. Diamantoulakis and G. K. Karagiannidis are with the Department of Electrical and Computer Engineering, Aristotle University of Thessaloniki, 54124 Thessaloniki, Greece (e-mails: toikonom@ece.auth.gr, tyrovolas@auth.gr, tegosoti@auth.gr, padiaman@auth.gr, geokarag@auth.gr).}
\thanks{P. Sarigiannidis is with the Department of Electrical and Computer Engineering, University of Western Macedonia, 50100 Kozani, Greece (e-mail: psarigiannidis@uowm.gr).}
}
\maketitle

\begin{abstract} 
This paper presents a maximum-likelihood detection framework that jointly mitigates hardware (HW) impairments in both amplitude and phase. By modeling transceiver distortions as residual amplitude and phase noise, we introduce the approximate phase-and-amplitude distortion detector (PAD-D), which operates in the polar domain and effectively mitigates both distortion components through distortion-aware weighting. The proposed detector performs reliable detection under generalized HW impairment conditions, achieving substantial performance gains over the conventional Euclidean detector (EUC-D) and the Gaussian-assumption phase noise detector (GAP-D), which is primarily designed to address phase distortions. In addition, we derive a closed-form high-SNR symbol error probability (SEP) approximation, which offers a generic analytical expression applicable to arbitrary constellations. Simulation results demonstrate that the PAD-D achieves up to an order-of-magnitude reduction in the error floor relative to EUC-D and GAP-D for both high-order quadrature amplitude modulation (QAM) and super amplitude phase-shift keying (SAPSK) constellations, establishing a unified and practical framework for detection under realistic transceiver impairments. Building on this framework, we further develop optimized constellations tailored to PAD-D, where the symbol positions are optimized in the complex plane to minimize SEP. The optimality of these constellations is confirmed through extensive simulations, which also verify the accuracy of the proposed analytical SEP approximation, even for the optimized designs.
\end{abstract}
\begin{IEEEkeywords}
Maximum likelihood detection, hardware impairments, symbol error-probability, constellation design
\end{IEEEkeywords}

\section{Introduction}
The increasing popularity of data-intensive applications, such as immersive extended reality, holographic telepresence, interactive cloud gaming, and high-definition visual streaming, has generated demand for next-generation wireless systems capable of delivering ultra-high data rates, reliability, and energy efficiency \cite{karag_survey, survey_v1, survey_v2}. To meet these requirements, a substantial increase in spectral efficiency is necessary, which is primarily achieved through high-order modulation formats that enable multiple bits to be transmitted per symbol \cite{high_order_constellations}. To sustain such operation, communication systems must ensure sufficiently high signal-to-noise ratio (SNR) levels to preserve detection reliability at these dense constellation configurations \cite{sapsk}. Meeting this requirement has driven major advances in physical-layer design, with technologies such as massive multiple-input multiple-output (MIMO) and reconfigurable intelligent surfaces (RIS) playing a central role in enhancing the effective SNR, improving channel conditioning, and enabling the practical deployment of high-order constellations without compromising link robustness \cite{heath_mimo, ris_hqam, zeris}. However, the densification of signal space introduces an intrinsic vulnerability to hardware (HW) impairments, since even small deviations in the amplitude or phase can distort symbol placement within the constellation, severely degrading overall performance. This limitation has been clearly observed in recent Wi-Fi standards, where the move to 4K-QAM in Wi-Fi 7 exposed the difficulty of maintaining reliability under generalized HW impairments. In Wi-Fi 8, alleviating such impairments is essential to enable higher modulation orders and achieve further throughput gains \cite{wifi-8}.

In practical transceivers, such imperfections arise from non-ideal components, such as oscillators, mixers, and power amplifiers (PAs), which introduce distortions that alter the amplitude and phase of the transmitted and received symbols. 
Among the most prominent sources are carrier frequency offset (CFO), oscillator phase noise (PN), amplifier nonlinearities, and in-phase quadrature (I/Q) imbalance, each contributing to deviations that can substantially degrade system performance \cite{cfo, phase_noise, pa_efficiency, iq_imbalances, rf-impairments}. Although modern calibration and digital compensation techniques can mitigate these effects to a certain extent, they cannot completely eliminate them, as impairments inherently exhibit stochastic behavior and vary dynamically with the instantaneous operating conditions of the HW \cite{dpd_residual_model, iq_imbalance_compensation, pn_compensation}. As a result, residual amplitude distortions (AD) and PN remain after compensation, resulting in a mismatch between the distortion-impaired received signal and the idealized additive white Gaussian noise (AWGN) model assumed by conventional detectors. This highlights the need for receiver designs that explicitly account for the statistical properties of residual AD and PN to ensure accurate detection and consistent reliability under realistic HW impairments.

\subsection{State-of-the-Art}
Extensive research efforts have been devoted to the characterization and mitigation of residual PN in modern communication systems, as it represents one of the dominant sources of transceiver distortion in high-order modulations. In \cite{foschini}, the residual PN was modeled as a Gaussian random process, commonly referred to as Gaussian PN (GPN), providing a tractable analytical framework for quantifying its statistical behavior and impact on symbol detection accuracy. This modeling approach enabled the derivation of closed-form performance metrics and laid the foundation for distortion-aware receiver design. Building on this theoretical basis, the authors in \cite{soft_metrics} proposed an approximate maximum-likelihood detection rule, termed the Gaussian-assumption phase noise detector (GAP-D), which explicitly incorporates the GPN statistics into the decision process, thus improving the receiver resilience against random phase fluctuations. 

More recently, several studies \cite{pqam, apsk_opt, eriksson_con_opt, dnn_con_opt_pn} extended this line of research toward constellation design and optimization under GPN-impaired conditions. These works explored both structured amplitude-and-phase shift keying (APSK) constellations and unstructured symbol configurations in the complex plane, demonstrating that tailoring the constellation geometry to the underlying PN distribution considerably improves symbol error probability (SEP) compared to conventional uniformly spaced constellations. However, despite the significant progress achieved in mitigating residual PN, these approaches remain inherently confined to the GPN framework and do not account for the more general case of HW impairments, where AD coexist with PN and jointly affect amplitude and phase statistics of the received signal, ultimately limiting detection performance.

Another major source of transceiver degradation arises from residual amplitude distortion (AD), which persists as a consequence of various HW impairments, most notably the nonlinear behavior of PAs, even after digital predistortion compensation. In high-order modulation formats, even small residual amplitude deviations can significantly alter the effective energy spacing between neighboring symbols, leading to increased detection errors. To mitigate these effects, several studies have focused on constellation optimization under amplitude distortion. In particular, \cite{spiral_nonlinear_channels} proposed a geometric shaping strategy that adapts the constellation to nonlinear transmission characteristics, improving robustness without explicitly modeling PN. Similarly, \cite{papr_constraint, papr_constraint_v2, papr_constraint_v3} investigated constellation design under peak-to-average power ratio (PAPR) constraints, aiming to limit the impact of PA nonlinearities by maintaining quasi-linear amplifier operation. More recently, the work in \cite{high_order_pn_ad} jointly considered PN and deterministic amplitude distortion resulting from PA nonlinearities, employing an autoencoder-based transmitter design to learn constellation mappings optimized for nonlinear channels. Despite these advancements, existing approaches largely rely on deterministic PA models or indirect PAPR control and do not capture the stochastic residual amplitude fluctuations that remain after practical calibration.

\subsection{Motivation \& Contribution}
Although the aforementioned studies address specific distortion mechanisms, they remain limited in scope, as they treat amplitude and phase impairments in isolation or rely on simplified deterministic models. In practice, however, transceiver distortion is inherently cumulative and stochastic, arising from the superposition of multiple stochastic HW non-idealities that jointly perturb both the amplitude and phase of transmitted symbols. Even after calibration and compensation, these effects persist as residual generalized HW impairments, which cannot be adequately represented by single-source models such as GPN or deterministic PA nonlinearities. This gap motivates the development of a novel detection framework capable of jointly characterizing and mitigating residual AD and PN, marking a significant step toward reliable high-order modulation under realistic transceiver conditions. To the best of the authors’ knowledge, no prior work has derived a maximum-likelihood detection rule that accounts for the combined statistical effects of residual AD and PN, nor formulated a constellation optimization framework that adapts symbol geometry to such generalized distortion conditions.

In this paper, we present an approximate maximum-likelihood detection framework, termed the phase-and-amplitude distortion detector (PAD-D), which jointly mitigates residual AD and PN arising from practical transceiver HW impairments. Specifically, the contributions of this work are summarized as follows:
\begin{itemize}
    \item We derive a polar-domain detector that incorporates the statistical characteristics of residual amplitude and phase distortions into the decision metric. The proposed PAD-D generalizes the GAP-D by introducing an additional amplitude-distortion term, effectively extending its applicability to generalized HW impairment conditions.
    \item We derive a closed-form high-SNR approximation for the SEP applicable to arbitrary constellation geometries and modulation orders. This analytical formulation provides tractable performance evaluation and offers insight into how residual AD and PN jointly affect the overall system performance.
    \item We formulate and solve a constellation optimization problem that minimizes the average SEP under joint amplitude and phase distortion conditions. The optimization is carried out using a hybrid global–local approach, combining simulated annealing for global exploration and gradient-based refinement for fine convergence under energy constraints. The resulting constellations exhibit distinct geometric adaptations that enhance robustness to generalized HW impairments.
    \item Extensive simulations demonstrate that PAD-D significantly outperforms both the GAP-D and the conventional Euclidean distance detector (EUC-D), reducing the error floor by up to an order of magnitude under generalized HW impairment conditions. Moreover, the analytical high-SNR SEP approximation remains accurate even for the optimized constellation designs, confirming its tightness and reliability as an analytical performance metric across different HW impairment regimes, while the optimized constellations derived under PAD-D further enhance robustness and deliver additional SEP gains compared to their conventional counterparts.
\end{itemize}

\subsection{Structure}
The remainder of the paper is organized as follows. Section II introduces the system model and the generalized impairment framework, encompassing residual amplitude distortion and PN. Section III develops the approximate maximum-likelihood detection metric for this model and derives a high-SNR SEP approximation applicable to arbitrary constellations. Section IV formulates the constellation optimization problem and details the hybrid simulated annealing with gradient refinement strategy used to obtain unstructured designs. Section V presents numerical results, including comparisons with benchmark detectors and validation of the analytical SEP. Section VI concludes the paper and outlines directions for future work.

\section{System Model and Preliminaries} \label{SM}
We consider a communication system that transmits a symbol $s\in\mathbb{C}$ that belongs to an $M$-ary constellation denoted by $\mathcal{C}$, with average energy $E_s$. By taking into account an AWGN channel impaired by both PN and residual AD, the received signal $r\in\mathbb{C}$ can be expressed as 
\begin{equation} \label{eq:system_model}
    \begin{aligned}
        r = g|s|e^{j\left(\phi + \arg\{s\}\right)} + n,
    \end{aligned}
\end{equation}
where $|\cdot|$ and $\arg\{\cdot\}$ denote the amplitude and the phase of a complex number, respectively, and $n$ represents the AWGN and is a complex Gaussian random variable with zero mean and variance $\sigma_n^2$. 
Moreover, the factor $g$ denotes the residual AD, i.e., a small multiplicative gain error that remains after amplitude-related compensation or calibration due to factors such as I/Q imbalances, PA nonlinearities, or AGC imperfections, and is modeled as $g \sim \mathcal{N}(1,\sigma_g^2)$, consistent with prior works where residual errors are well approximated as Gaussian \cite{dpd_residual_model}. Furthermore, $\phi$ denotes the PN resulting from imperfect phase recovery, which has traditionally been modeled as a Tikhonov-distributed random variable with zero mean and variance $\sigma_{\phi}^2$ \cite{proakis2002communication}. However, in modern communication systems, where $\pn^2 \ll 1$, it can be accurately represented as a Gaussian PN (GPN) with zero mean and variance $\pn^2$ \cite{akyldiz}. Finally, practical scenarios are typically classified according to the PN level so that strong, moderate, and low PN conditions correspond to $\pn^2 = 10^{-1}$, $\pn^2 = 10^{-2}$, and $\pn^2 = 10^{-4}$, respectively \cite{bicais_pn_levels}.

Typically, in digital communication systems where GPN and AD are neglected, symbol detection is performed using the EUC-D in the I/Q plane, which is designed for the AWGN channel and operates by selecting the constellation point with the minimum Euclidean distance to the received symbol $r$, such that the detected symbol is given by
\begin{equation}\label{EUC-D}
    \begin{aligned}
        \hat{s} = \arg \min_{s\in\mathcal{C}}\hspace{4px}\lvert r-s\rvert^2.
    \end{aligned}
\end{equation}
However, in the presence of GPN, the optimal maximum-likelihood detection strategy is no longer equivalent to EUC-D. In this direction, a detector tailored for GPN-impaired channels, denoted as GAP-D, was derived in \cite{soft_metrics}, and is expressed as
\begin{equation}\label{GAP-D}
    \begin{aligned}
    \hat{s} = \arg \min_{s\in\mathcal{C}}\hspace{4px}&\frac{\big(\lvert r \rvert-\lvert s \rvert \big)^2}{\frac{\sigma_{n}^2}{2}} + \frac{\big(\arg\{r\}-\arg\{s\}\big)^2}{\sigma_{\phi}^2 + \frac{\sigma_{n}^2}{2\lvert s \rvert^2}} \\
    &+ \log\left(\sigma_{\phi}^2 + \frac{\sigma_{n}^2}{2\lvert s \rvert^2}\right).
\end{aligned}
\end{equation}
As observed in \eqref{GAP-D}, the performance of GAP-D is jointly determined by $\sigma_n^2$ and $\sigma_\phi^2$, while the decision metric explicitly incorporates both the amplitude and phase of the received symbol $r$ relative to the constellation point $s$. This formulation highlights that GAP-D operates in the polar domain, where the impact of AWGN and GPN is jointly mitigated.

\begin{figure}[h!]
	\centering
	\begin{tikzpicture}
	\begin{semilogyaxis}[
    width=0.9\linewidth,
	xlabel = $\frac{E_{s}}{\sigma_n^2}$ (dB),
	ylabel = Symbol Error Probability (SEP),
	xmin = 0,
	xmax = 60,
	ymin = 0.00001,
	ymax = 1,
    xtick = {0,10,...,80},
	grid = major,
	legend entries = {{32-QAM (EUC-D)}, {32-QAM (GAP-D)}, {256-QAM (EUC-D)}, {256-QAM (GAP-D)}},
	legend cell align = {left},
    legend pos = south east,
    legend style={font=\tiny}
	]
	
	\addplot[
        color = black,
        no marks,
	mark repeat = 1,
	mark size = 3,
        line width = 1pt,
        style = solid
	]
	table {figures/simulation_results/qam_32_eucd_sigma_g_1e-3_sigma_phi_1e-4.txt};
        \addplot[
        color = black,
	mark = diamond*,
	mark repeat = 1,
	mark size = 3,
        line width = 1pt
	]
	table {figures/simulation_results/qam_32_gapd_sigma_g_1e-3_sigma_phi_1e-4.txt};
        \addplot[
        color = red,
	no marks,
	mark repeat = 1,
	mark size = 2,
        style=solid,
        line width = 1pt
	]
	table {figures/simulation_results/qam_256_eucd_sigma_g_1e-3_sigma_phi_1e-4.txt};
     \addplot[
        color = red,
	mark=square*,
	mark repeat = 1,
	mark size = 2,
        style=solid,
        line width = 1pt
    ]
	table {figures/simulation_results/qam_256_gapd_sigma_g_1e-3_sigma_phi_1e-4.txt};
	\end{semilogyaxis}
	\end{tikzpicture}
	\caption{EUC-D vs GAP-D under $\sigma_g^2=10^{-3}$ and  $\sigma^2_{\phi} = 10^{-4}$}
	\label{fig:euc_vs_gap}
\end{figure}
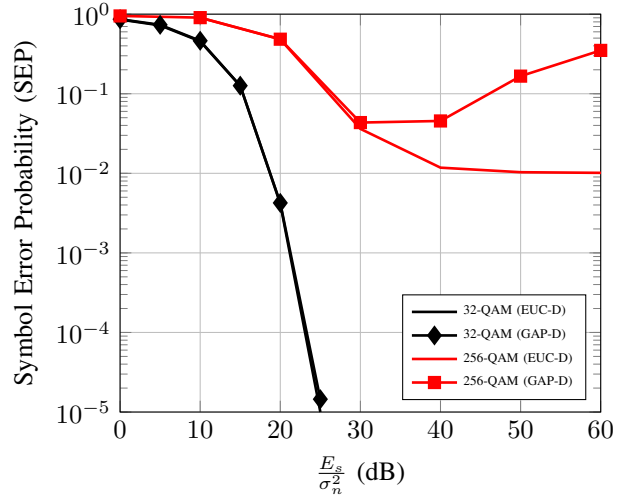

However, as illustrated in Fig. 1, both the EUC-D and GAP-D exhibit inherent limitations when HW impairments become non-negligible. While their performance remains comparable under small AD and PN variances for low-order constellations such as 32-QAM, where the effect of these residual impairments on the decision boundaries is minor and both detectors can still make near-optimal decisions, the situation changes significantly for higher-order constellations. In more detail, as the constellation order increases, the impact of even small HW impairments becomes increasingly visible, and both detectors fail to fully compensate for the combined effects of AD and PN, resulting in the emergence of a clear error floor at high SNR. In particular, GAP-D accounts for AD only through its dependence on the AWGN component, which vanishes as SNR increases, leaving the detector effectively insensitive to residual amplitude-related distortions. Consequently, its polar-domain weighting overemphasizes the PN contribution, which compresses the symbol rings in the complex plane and increases inter-ring interference. This behavior leads to a degradation in SEP as the SNR rises, underscoring the need for a more comprehensive maximum-likelihood formulation that jointly accounts for both residual AD and PN to suppress the error floor and achieve reliable detection under generalized HW impairment conditions.

\section{Detector Design for Generalized HW Impairments}
In this section, an approximate maximum-likelihood detection rule is developed for the generalized system model in \eqref{eq:system_model}, which jointly accounts for both amplitude and phase distortions introduced by the transceiver HW. The resulting detector, referred to as the PAD-D, operates in the polar domain and provides a unified framework for mitigating the combined effects of AD and PN. Building on this metric, we further derive a closed-form SEP approximation applicable to arbitrary constellations, enabling analytical performance evaluation without the need for extensive simulations.
\subsection{Detection Metric}
In this subsection, we derive an approximate detection metric for the generic system model described in \eqref{eq:system_model}. The maximum-likelihood detection rule is formally given by 
\begin{equation}
    \begin{aligned}
        \hat{s} = \arg \max_{s\in\mathcal{C}}\hspace{4px} p(r|s),
    \end{aligned}
\end{equation}
with the likelihood function expressed as
\begin{equation}\label{eq:conditional_prob_ML}
    \begin{aligned}
        p(r|s) = \int_{-\pi}^{\pi}\int_{0}^{\infty}p(r|s,\phi,g)p(\phi)p(g)dgd\phi,
    \end{aligned}
\end{equation}
where $p(\phi)$ and $p(g)$ denote the marginal probability density functions of the PN and amplitude, respectively, and $g$ is modeled as $g\sim\mathcal{N}\left(1,\sigma_g^2\right)$, while the integral is formally restricted to $g\geq0$. For small values of $\sigma_g^2 \ll 1$, which are typical in practical scenarios, the probability of drawing negative gains is negligible, and the Gaussian model with this truncation provides an accurate and tractable representation of residual gain fluctuations. However, the double integral in \eqref{eq:conditional_prob_ML} does not admit a closed-form solution, and thus, the exact maximum-likelihood detection rule is not analytically tractable. To overcome this issue, we derive an approximate metric by processing the received symbol and the constellation symbols in their polar form. Specifically, the amplitude and phase decision residuals, i.e., the differences between the received sample’s magnitude and phase and those of each constellation point, are modeled as Gaussian-distributed variables whose effective variances depend on both additive noise and HW impairments. This approximation enables us to replace the intractable integral with a closed-form detection rule, which generalizes the GAP-D framework to jointly account for AD and PN. The proposed metric is presented in the following theorem.
\begin{theorem}\label{prop:detection_metric}
The transmitted symbol $s \in \mathcal{C}$, when impaired by AWGN, GPN, and residual AD, can be reliably detected using the PAD-D, which is given by
\begin{equation}\label{eq:proposed_metric}
    \begin{aligned}
        \hat{s} = \arg \min_{s_m\in\mathcal{C}}\hspace{4px} \mathcal{L}(s_m),
    \end{aligned}
\end{equation}
where 
\begin{equation}\label{eq:L}
    \begin{aligned}
        \mathcal{L}(s_m) = &\frac{\left(|r|-|s|\right)^2}{V_{m}^{(a)}} +\frac{\left(\arg\{r\}-\arg\{s\}\right)^2}{V_{m}^{(\theta)}} \\
         &+\ln\left(V_{m}^{(a)}V_{m}^{(\theta)}\right),
    \end{aligned}
\end{equation}
with $V_{m}^{(a)} =\sigma_n^2/2+\sigma_g^2|s_m|^2$ and $V_{m}^{(\theta)} = \pn^2 + \sigma_n^2/(2|s_m|^2)$.
\end{theorem}
\begin{IEEEproof}
The proof is provided in Appendix A.
\end{IEEEproof}

\begin{remark}
The proposed PAD-D metric reduces to the GAP-D when $\sigma_g^2\rightarrow 0$, thus extending GAP-D to account for AD and PN jointly.
\end{remark}

The PAD-D detection rule can be interpreted as operating in the polar plane. The first term in \eqref{eq:proposed_metric} penalizes residuals along the amplitude axis, with a normalization that accounts for both additive noise and the variance of AD, reflecting HW-induced fluctuations in symbol magnitude. The second term in \eqref{eq:proposed_metric} penalizes residuals along the angular axis, normalized by the combined effect of additive noise and PN variance, thus compensating for random phase rotations. Through this polar-domain formulation, the detector adaptively balances the contributions of amplitude and phase decision residuals, effectively mitigating distortions introduced by HW impairments in both amplitude and phase.

\subsection{Performance Analysis}
The analysis is based on the union-bound framework, where the SEP is approximated by averaging the pairwise error probabilities between distinct constellation symbols under the high-SNR assumption. In more detail, the pairwise error probability of the PAD-D is obtained by applying moment matching to the amplitude and phase decision residual terms in the decision metric, yielding a tractable characterization of the SEP performance under joint AD and PN.

Consider two distinct constellation symbols $s_i$ and $s_j$, where a pairwise error event corresponds to the case in which $s_i$ is transmitted but the detector selects $s_j$ instead. For the PAD-D, this probability can be expressed as 
\begin{equation}
    \begin{aligned}
        P\left(s_i\rightarrow s_j|s_i\right) = P\left(\mathcal{L}_j-\mathcal{L}_i<0|s_i\right),      
    \end{aligned}
\end{equation}
where $\mathcal{L}_k = \mathcal{L}(s_k)$. In the following theorem, we provide an analytical approximation of the SEP for arbitrary constellations.
\begin{theorem}\label{prop:SEP_approx}
    The SEP for high-order constellations under generalized HW impairments can be approximated as
    \begin{equation}\label{eq:SEP_approx}
    \small
        \begin{aligned}
            P_e \approx \frac{1}{M}\sum_{i=1}^{M}\sum_{j\neq i} \left(Q\left(\frac{\xi_{ij}}{\omega_{ij}}\right) - 2T\left(\frac{-\xi_{ij}}{\omega_{ij}},\frac{\delta_{ij}}{\sqrt{1-\delta_{ij}}}\right)\right),   \end{aligned}
    \end{equation}
    where $Q(\cdot)$ is the Gaussian $Q$-function \cite{karag_Q}, $T(\cdot,\cdot)$ is the Owen's $T$ function \cite{abramowitz}, and $\omega_{ij}$, $\xi_{ij}$ are given by \eqref{eq:omega} and \eqref{eq:xi}, respectively, at the top of the next page. Moreover, it holds that
    \begin{equation}\label{eq:delta_ij}
        \begin{aligned}
            \delta_{ij}
= \operatorname{sgn}(\gamma_{1,ij})
\sqrt{\frac{\left(\frac{2\pi^{3/2}|\gamma_{1,ij}|}{2^{3/2}(4-\pi)}\right)^{2/3}}{1+\frac{2}{\pi}\left(\frac{2\pi^{3/2}|\gamma_{1,ij}|}{2^{3/2}(4-\pi)}\right)^{2/3}}},
        \end{aligned}
    \end{equation}
    where $\operatorname{sgn}(\cdot)$ is the sign function \cite{abramowitz}, and $\gamma_{1,ij}$ is given by \eqref{eq:skewness} at the top of the next page.
\end{theorem}
\begin{IEEEproof}
    The proof is provided in Appendix B.
\end{IEEEproof}
\begin{figure*}
    \begin{equation}\label{eq:omega}
       \begin{aligned}
            \omega_{ij}&=\sqrt{\frac{2\left(\frac{1}{V_j^{(a)}} - \frac{1}{V_i^{(a)}}\right)^2\left(V_i^{(a)}\right)^2 + 4\left(\frac{|s_j|-|s_i|}{V_j^{(a)}}\right)^2V_i^{(a)} + 4\left(\frac{\arg(s_j)-\arg(s_i)}{V_j^{(\theta)}}\right)^2V_i^{(\theta)}}{1-2\delta_{ij}^2/\pi}}
        \end{aligned}
    \end{equation}
    \begin{equation}\label{eq:xi}
       \begin{aligned}
            \xi_{ij}&=\frac{V_{i}^{(a)}}{V_{j}^{(a)}}-1 + \frac{(|s_j|-|s_i|)^2}{V_j^{(a)}} 
            + \frac{(\arg(s_j)-\arg(s_i))^2}{V_j^{(\theta)}} 
            + \log\left(\frac{V_j^{(a)} V_j^{(\theta)}}{V_i^{(a)} V_i^{(\theta)}}\right) - \omega_{ij}\delta_{ij}\sqrt{\frac{2}{\pi}}
        \end{aligned}
    \end{equation}
    \begin{equation}\label{eq:skewness}
       \begin{aligned}
            \gamma_{1,ij}&=\frac{8\left(\frac{V_i^{(a)}}{V_j^{(a)}} - 1\right)^3 + 24\left(\frac{1}{V_j^{(a)}} - \frac{1}{V_i^{(a)}}\right)\left(\frac{|s_j|-|s_i|}{V_j^{(a)}}\right)^2\left(V_i^{(a)}\right)^2}{\left(2\left(\frac{1}{V_j^{(a)}} - \frac{1}{V_i^{(a)}}\right)^2\left(V_i^{(a)}\right)^2 + 4\left(\frac{|s_j|-|s_i|}{V_j^{(a)}}\right)^2V_i^{(a)} + 4\left(\frac{\arg(s_j)-\arg(s_i)}{V_j^{(\theta)}}\right)^2V_i^{(\theta)}\right)^{3/2}}
        \end{aligned}
    \end{equation}
    \hrule
\end{figure*}
\begin{remark}
    By setting $\sigma_n^2\rightarrow0$ in \eqref{eq:SEP_approx}, the parameters in \eqref{eq:omega}-\eqref{eq:skewness} become functions only of the AD variance $\sigma_g^2$, the PN variance $\pn^2$, and the constellation geometry. In this regime, the analytical SEP approximation simplifies to a closed-form expression that quantifies the irreducible error floor imposed by the residual HW impairments. This closed-form characterization provides valuable insight into the ultimate performance limit of the system, revealing how amplitude and phase distortions jointly constrain detection accuracy even in the absence of additive noise.
\end{remark}
\section{Constellation Optimization}
In this section, we present the formulation of the constellation optimization problem, whose objective is the minimization of the average SEP under the proposed PAD-D detection metric. To this end, we derive novel unstructured constellations that are inherently adapted to the characteristics of the PAD-D receiver and the underlying HW impairments. In this framework, the position of each symbol is freely optimized in the complex plane through a global search method that combines simulated annealing with a subsequent gradient-based refinement, allowing effective global exploration while ensuring accurate convergence to high-quality solutions. This hybrid approach provides full flexibility in shaping the constellation geometry and achieves near-optimal designs despite the inherently non-convex nature of the problem.

Formally, let $\boldsymbol{s} = [s_1, s_2, \dots, s_M]^T$ denote the complex-valued constellation symbols. The optimization problem is expressed as
\begin{equation*}\tag{\textbf{P1}}\label{eq:maxmin_with_theta}
\begin{aligned}
    \begin{array}{cl}
    &\mathop{\mathrm{min}}\limits_{\boldsymbol{s}}\quad
    P_e(\boldsymbol{s}) \\
    &\text{s.t.}
    \quad\mathrm{C}_1: \frac{1}{M}\sum_{m=1}^{M}|s_m|^2=1, \\
    & \quad \hspace{0.2em}\quad\mathrm{C}_2:\sum_{m=1}^{M}s_m = 0, \\
    \end{array}
\end{aligned}
\end{equation*}
where $P_e(\boldsymbol{s})$ represents the average SEP achieved by the PAD-D under residual amplitude and phase distortions. This formulation equally applies to the EUC-D and GAP-D, with $P_e(\boldsymbol{s})$ evaluated under their respective decision rules, allowing a unified comparison of the optimal constellation structures under generalized HW impairments. Although a closed-form SEP approximation was derived earlier, $P_e(\boldsymbol{s})$ is here evaluated numerically through Monte Carlo simulations to ensure reliable accuracy across the entire SNR range. For each candidate constellation, random realizations of the distortion parameters and additive noise are drawn from their respective distributions with fixed statistics to maintain consistent impairment conditions throughout the optimization process. Fixed random seeds and balanced symbol sampling are employed to produce a smooth and reproducible objective surface. The normalization constraints enforce unit average symbol power and ensure that the optimized constellations remain centered around the origin. This symmetry condition eliminates any unnecessary bias in the signal’s mean value, preventing power inefficiencies and distortion-dependent shifts in the effective constellation centroid.

Since the detection error probability depends jointly and nonlinearly on the symbol amplitudes, phases, and their interaction with additive noise, the objective landscape of $P_e(\boldsymbol{s})$ is highly non-convex. To handle this effectively, we adopt a hybrid global–local optimization framework implemented using MATLAB’s toolboxes. The global exploration stage employs the simulated annealing algorithm from the global optimization toolbox, which performs a probabilistic search over the feasible space while respecting the same normalization constraints imposed on the constellation. This allows simulated annealing to explore widely and occasionally accept higher-cost configurations to escape poor local minima and discover promising regions of the design space. Once a suitable region is identified, a gradient-based constrained optimization method from MATLAB’s optimization toolbox is used to perform local refinement under the same constraints, precisely adjusting the constellation points to achieve fine-grained performance improvement. By combining the complementary strengths of the two stages, simulated annealing delivers robust global exploration across a rough, non-convex landscape, while the gradient-based refinement guides the search to converge efficiently and with numerical stability in the near-optimal region. As a result, this approach achieves both exploration breadth and refinement accuracy, resulting in near-optimal constellations that balance robustness and efficiency under generalized HW impairments.

\begin{figure*}
\centering
\begin{minipage}[t]{0.3\textwidth}
\centering
\begin{tikzpicture}[scale=0.5]
\begin{axis}[
        xtick=\empty,
        ytick=\empty,
        xlabel={$\text{Quadrature Q}$},
        ylabel={$\text{In-Phase I}$},
        xlabel style={below},
        ylabel style={above},
        xmin=-2,
        xmax=2,
        ymin=-2,
        ymax=2,
    ]
    \addplot [only marks, mark size = 3, black] table {figures/data/IQ_opt_64_SNR_20dB_eucd_sigma_g_1e-2_sigma_phi_1e-3.txt};
\end{axis}
\end{tikzpicture}
\subcaption{EUC-D at SNR $20$ dB}
\end{minipage}
\begin{minipage}[t]{0.3\textwidth}
\centering
\begin{tikzpicture}[scale=0.5]
\begin{axis}[
        xtick=\empty,
        ytick=\empty,
        xlabel={$\text{Quadrature Q}$},
        ylabel={$\text{In-Phase I}$},
        xlabel style={below},
        ylabel style={above},
        xmin=-2,
        xmax=2,
        ymin=-2,
        ymax=2,
    ]
    \addplot [only marks, mark size = 3,black] table {figures/data/IQ_opt_64_SNR_20dB_gapd_sigma_g_1e-2_sigma_phi_1e-3.txt};
\end{axis}
\end{tikzpicture}
\subcaption{GAP-D at SNR $20$ dB}
\end{minipage}
\begin{minipage}[t]{0.3\textwidth}
\centering
\begin{tikzpicture}[scale=0.5]
\begin{axis}[
        xtick=\empty,
        ytick=\empty,
        xlabel={$\text{Quadrature Q}$},
        ylabel={$\text{In-Phase I}$},
        xlabel style={below},
        ylabel style={above},
        xmin=-2,
        xmax=2,
        ymin=-2,
        ymax=2,
    ]
    \addplot [only marks, mark size = 3, black] table {figures/data/IQ_opt_64_SNR_20dB_proposed_sigma_g_1e-2_sigma_phi_1e-3.txt};	
    \end{axis}
    \end{tikzpicture}   
    \subcaption{PAD-D at SNR $20$ dB}
\end{minipage}
\begin{minipage}[t]{0.3\textwidth}
\centering
\begin{tikzpicture}[scale=0.5]
\begin{axis}[
        xtick=\empty,
        ytick=\empty,
        xlabel={$\text{Quadrature Q}$},
        ylabel={$\text{In-Phase I}$},
        xlabel style={below},
        ylabel style={above},
        xmin=-2,
        xmax=2,
        ymin=-2,
        ymax=2,
    ]
    \addplot [only marks, mark size = 3, black] table {figures/data/IQ_opt_64_SNR_30dB_eucd_sigma_g_1e-2_sigma_phi_1e-3.txt};
\end{axis}
\end{tikzpicture}
\subcaption{EUC-D at SNR $30$ dB}
\end{minipage}
\begin{minipage}[t]{0.3\textwidth}
\centering
\begin{tikzpicture}[scale=0.5]
\begin{axis}[
        xtick=\empty,
        ytick=\empty,
        xlabel={$\text{Quadrature Q}$},
        ylabel={$\text{In-Phase I}$},
        xlabel style={below},
        ylabel style={above},
        xmin=-2,
        xmax=2,
        ymin=-2,
        ymax=2,
    ]
    \addplot [only marks, mark size = 3, black] table {figures/data/IQ_opt_64_SNR_30dB_gapd_sigma_g_1e-2_sigma_phi_1e-3.txt};
    \end{axis}
    \end{tikzpicture}  
    \subcaption{GAP-D at SNR $30$ dB}
\end{minipage}
\begin{minipage}[t]{0.3\textwidth}
\centering
\begin{tikzpicture}[scale=0.5]
\begin{axis}[
        xtick=\empty,
        ytick=\empty,
        xlabel={$\text{Quadrature Q}$},
        ylabel={$\text{In-Phase I}$},
        xlabel style={below},
        ylabel style={above},
        xmin=-2,
        xmax=2,
        ymin=-2,
        ymax=2,
    ]
    \addplot [only marks, mark size = 3, black] table {figures/data/IQ_opt_64_SNR_30dB_proposed_sigma_g_1e-2_sigma_phi_1e-3.txt};
\end{axis}
\end{tikzpicture}
\subcaption{PAD-D at SNR $30$ dB}
\end{minipage}
\caption{Optimized 64-point constellations for EUC-D, GAP-D, and PAD-D under $\sigma_g^2=10^{-2}$ and $\pn^2 = 10^{-3}$.}
\label{fig:optimized_scatterplots}
\end{figure*}
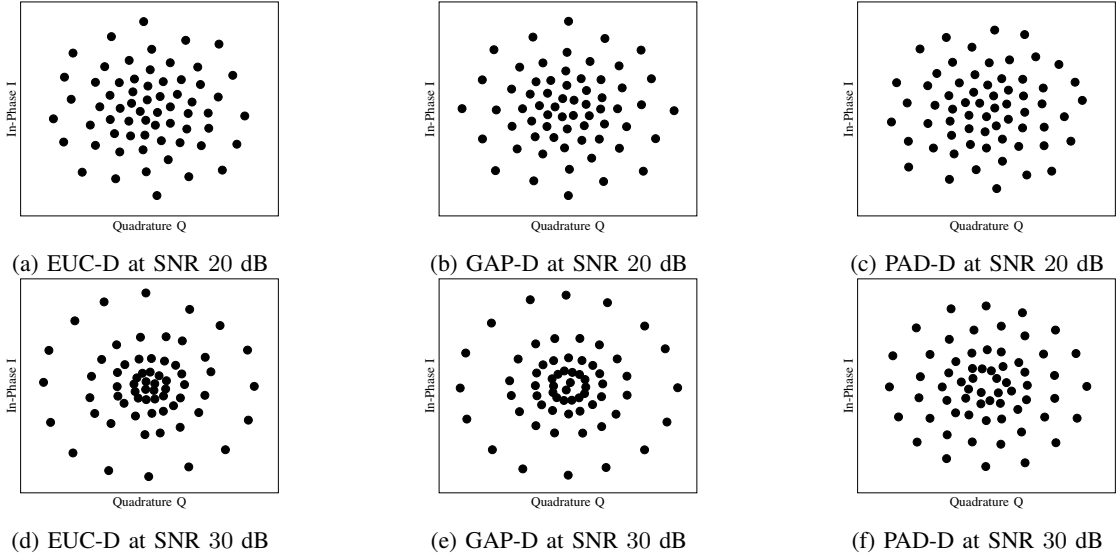

To provide additional intuition on how this optimization process shapes the constellation geometry under realistic transceiver distortions, Fig. \ref{fig:optimized_scatterplots} illustrates representative optimized constellations obtained for different detectors and SNR conditions. These results correspond to $M = 64$ optimized constellations under $\sigma_g^2=10^{-2}$ and $\pn^2=10^{-3}$, representing a moderate distortion regime in which both AD and PN jointly influence detection performance. In more detail, at 20 dB, the constellations optimized for the EUC-D and GAP-D already begin to deviate from the ideal QAM geometry, as the optimization process reacts to the presence of residual AD and PN without the benefit of a distortion-aware metric. As the EUC-D models all impairments as additive Gaussian noise, it cannot capture the statistical dependence between symbol amplitude, phase, and the distortion processes. Therefore, it compensates geometrically by expanding the constellation in the radial direction. While this increases the distance between rings and partially mitigates overlap caused by amplitude fluctuations, it also reduces angular compactness and leads to inefficient power allocation. The GAP-D, on the other hand, assumes only PN and neglects the presence of AD. This results in an even stronger radial expansion, as its likelihood formulation implicitly presumes amplitude stability. Consequently, the optimization increases the ring spacing to preserve angular separability, but this comes at the cost of heightened sensitivity to amplitude variations. As the SNR increases to 30 dB, where additive noise becomes negligible and residual distortions dominate, the constellations optimized using these two detectors spread out more, revealing their reliance on geometric expansion rather than true distortion adaptation. In contrast, constellations optimized for the proposed PAD-D remain compact and balanced at both SNR levels since the PAD-D explicitly incorporates joint AD and PN statistics in its detection metric. Consequently, it can compensate for their effects analytically without excessive radial spacing. This results in constellations characterized by smaller inter-ring distances, smoother angular symmetry, and more uniform energy distribution, reflecting an efficient balance between symbol distinguishability and energy utilization. Overall, PAD-D optimization achieves robustness through model-based adaptation rather than geometric separation heuristics, producing constellation structures that are resilient to both AD and PN across different SNR regimes.

\section{Simulation Results}
In this section, we present numerical results to evaluate the performance of the proposed system and validate the analytical derivations. The numerical results were obtained through Monte Carlo simulations involving $10^{7}$ transmitted symbols to ensure statistical reliability. To assess the generality of the proposed detector and its compatibility with practical modulation formats, we consider both conventional quadrature amplitude modulation (QAM) constellation, which is widely adopted in modern communication standards, and super-APSK (SAPSK), where a SAPSK$(M,\Gamma)$ is a structured APSK constellation where $\Gamma$ represents the number of energy levels, jointly defining both the energy levels and the number of symbols per level, thus reducing the optimization to a single dimension \cite{sapsk}. This property renders SAPSK highly flexible, enabling efficient SNR-dependent adaptation and increased robustness to HW impairments through its inherent structural degree of freedom.

\begin{figure}
	\centering
    \begin{subfigure}{\linewidth}
    \centering
    \begin{tikzpicture}
	\begin{semilogyaxis}[
    width=0.95\linewidth,
	xlabel = $\frac{E_{s}}{\sigma_n^2}$ (dB),
	ylabel = Symbol Error Probability (SEP),
	xmin = 10,
	xmax = 80,
	ymin = 0.00001,
	ymax = 1,
    xtick = {10, 20,...,80},
	grid = major,
	legend cell align = {left},
    legend pos = south west,
    legend style={font=\footnotesize}
	]

	\addplot[
	no marks,
    color=black,
    line width = 1pt,
	style = dashdotted,
	]
	table {figures/simulation_results/qam_256_eucd_sigma_g_1e-4_sigma_phi_1e-3.txt};
    \addlegendentry{EUC-D};
    
    \addplot[
	no marks,
    color=black,
    line width = 1pt,
	style = dashed,
	]
	table {figures/simulation_results/qam_256_gapd_sigma_g_1e-4_sigma_phi_1e-3.txt};
    \addlegendentry{GAP-D};
    
    \addplot[
	no marks,
    color=black,
    line width = 1pt,
	style = solid,
	]
	table {figures/simulation_results/qam_256_proposed_sigma_g_1e-4_sigma_phi_1e-3.txt};
    \addlegendentry{PAD-D};

     \addplot[
	only marks,
    mark=square,
    mark options = solid,
    mark repeat = 1,
    mark size = 2.8,
    color=red,
    line width = 1pt,
	style = dashdotted,
	]
	table {figures/simulation_results/qam_128_eucd_sigma_g_1e-4_sigma_phi_1e-3.txt};
    \addlegendentry{128-QAM};
    
    \addplot[
	only marks,
    mark=o,
    mark options = solid,
    mark repeat = 1,
    mark size = 2.8,
    color=black,
    line width = 1pt,
	style = dashdotted,
	]
	table {figures/simulation_results/qam_256_eucd_sigma_g_1e-4_sigma_phi_1e-3.txt};
    \addlegendentry{256-QAM};

    \addplot[
	only marks,
    mark=o,
    mark options = solid,
    mark repeat = 1,
    mark size = 2.8,
    color=black,
    line width = 1pt,
	style = dashdotted,
	]
	table {figures/simulation_results/qam_256_gapd_sigma_g_1e-4_sigma_phi_1e-3.txt};

    \addplot[
	only marks,
    mark=o,
    mark options = solid,
    mark repeat = 1,
    mark size = 2.8,
    color=black,
    line width = 1pt,
	style = dashdotted,
	]
	table {figures/simulation_results/qam_256_proposed_sigma_g_1e-4_sigma_phi_1e-3.txt};
    
    \addplot[
	only marks,
    mark=square,
    mark options = solid,
    mark repeat = 1,
    mark size = 2.8,
    color=red,
    line width = 1pt,
	style = dashed,
	]
	table {figures/simulation_results/qam_128_gapd_sigma_g_1e-4_sigma_phi_1e-3.txt};

    \addplot[
	only marks,
    mark=square,
    mark options = solid,
    mark repeat = 1,
    mark size = 2.8,
    color=red,
    line width = 1pt,
	style = solid,
	]
	table {figures/simulation_results/qam_128_proposed_sigma_g_1e-4_sigma_phi_1e-3.txt};

    \addplot[
	no marks,
    color=red,
    line width = 1pt,
	style = dashdotted,
	]
	table {figures/simulation_results/qam_128_eucd_sigma_g_1e-4_sigma_phi_1e-3.txt};

    \addplot[
	no marks,
    color=red,
    line width = 1pt,
	style = dashed,
	]
	table {figures/simulation_results/qam_128_gapd_sigma_g_1e-4_sigma_phi_1e-3.txt};

    \addplot[
	no marks,
    color=red,
    line width = 1pt,
	style = solid,
	]
	table {figures/simulation_results/qam_128_proposed_sigma_g_1e-4_sigma_phi_1e-3.txt};

	\end{semilogyaxis}
	\end{tikzpicture}
	\subcaption{$\sigma_g^2=10^{-4}$ and $\pn^2=10^{-3}$}
    \end{subfigure}

    \begin{subfigure}{\linewidth}
    \centering
    \begin{tikzpicture}
	\begin{semilogyaxis}[
    width=0.95\linewidth,
	xlabel = $\frac{E_{s}}{\sigma_n^2}$ (dB),
	ylabel = Symbol Error Probability (SEP),
	xmin = 10,
	xmax = 80,
	ymin = 0.00001,
	ymax = 1,
    xtick = {10, 20,...,80},
	grid = major,
	legend cell align = {left},
    legend pos = south west,
    legend style={font=\footnotesize}
	]

    	\addplot[
	no marks,
    color=black,
    line width = 1pt,
	style = dashdotted,
	]
	table {figures/simulation_results/qam_256_eucd_sigma_g_1e-3_sigma_phi_1e-4.txt};
    \addlegendentry{EUC-D};
    
    \addplot[
	no marks,
    color=black,
    line width = 1pt,
	style = dashed,
	]
	table {figures/simulation_results/qam_256_gapd_sigma_g_1e-3_sigma_phi_1e-4.txt};
    \addlegendentry{GAP-D};
    
    \addplot[
	no marks,
    color=black,
    line width = 1pt,
	style = solid,
	]
	table {figures/simulation_results/qam_256_proposed_sigma_g_1e-3_sigma_phi_1e-4.txt};
    \addlegendentry{PAD-D};

     \addplot[
	only marks,
    mark=square,
    mark options = solid,
    mark repeat = 1,
    mark size = 2.8,
    color=red,
    line width = 1pt,
	style = dashdotted,
	]
	table {figures/simulation_results/qam_128_eucd_sigma_g_1e-3_sigma_phi_1e-4.txt};
    \addlegendentry{128-QAM};
    
    \addplot[
	only marks,
    mark=o,
    mark options = solid,
    mark repeat = 1,
    mark size = 2.8,
    color=black,
    line width = 1pt,
	style = dashdotted,
	]
	table {figures/simulation_results/qam_256_eucd_sigma_g_1e-3_sigma_phi_1e-4.txt};
    \addlegendentry{256-QAM};

    \addplot[
	only marks,
    mark=o,
    mark options = solid,
    mark repeat = 1,
    mark size = 2.8,
    color=black,
    line width = 1pt,
	style = dashdotted,
	]
	table {figures/simulation_results/qam_256_gapd_sigma_g_1e-3_sigma_phi_1e-4.txt};

    \addplot[
	only marks,
    mark=o,
    mark options = solid,
    mark repeat = 1,
    mark size = 2.8,
    color=black,
    line width = 1pt,
	style = dashdotted,
	]
	table {figures/simulation_results/qam_256_proposed_sigma_g_1e-3_sigma_phi_1e-4.txt};
    
    \addplot[
	only marks,
    mark=square,
    mark options = solid,
    mark repeat = 1,
    mark size = 2.8,
    color=red,
    line width = 1pt,
	style = dashed,
	]
	table {figures/simulation_results/qam_128_gapd_sigma_g_1e-3_sigma_phi_1e-4.txt};

    \addplot[
	only marks,
    mark=square,
    mark options = solid,
    mark repeat = 1,
    mark size = 2.8,
    color=red,
    line width = 1pt,
	style = solid,
	]
	table {figures/simulation_results/qam_128_proposed_sigma_g_1e-3_sigma_phi_1e-4.txt};

    \addplot[
	no marks,
    color=red,
    line width = 1pt,
	style = dashdotted,
	]
	table {figures/simulation_results/qam_128_eucd_sigma_g_1e-3_sigma_phi_1e-4.txt};

    \addplot[
	no marks,
    color=red,
    line width = 1pt,
	style = dashed,
	]
	table {figures/simulation_results/qam_128_gapd_sigma_g_1e-3_sigma_phi_1e-4.txt};

    \addplot[
	no marks,
    color=red,
    line width = 1pt,
	style = solid,
	]
	table {figures/simulation_results/qam_128_proposed_sigma_g_1e-3_sigma_phi_1e-4.txt};
	\end{semilogyaxis}
	\end{tikzpicture}
	\subcaption{$\sigma_g^2=10^{-3}$ and $\pn^2=10^{-4}$}
    \end{subfigure}
    \caption{Comparison between EUC-D, GAP-D, and PAD-D for QAM.}
	\label{fig::detectors_comparison_qam}
\end{figure}
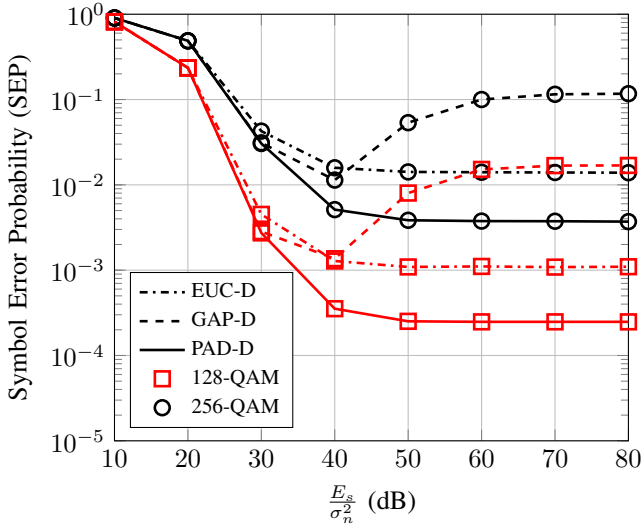
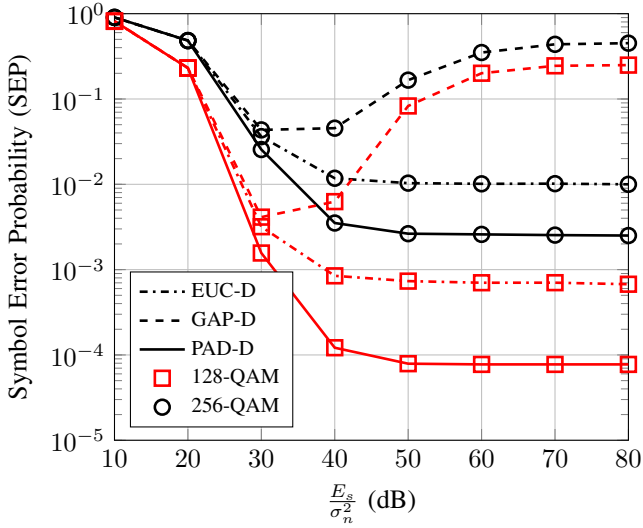

Fig. \ref{fig::detectors_comparison_qam} compares the performance of the conventional EUC-D, the GAP-D, and the proposed PAD-D for higher-order QAM constellations, i.e., $M=\{128, 256\}$. In Fig. \ref{fig::detectors_comparison_qam}a, the PN variance dominates with $\pn^2 > \sigma_g^2$, whereas Fig. \ref{fig::detectors_comparison_qam}b the AD variance dominates with $\pn^2 < \sigma_g^2$. Across both scenarios, the proposed PAD-D consistently achieves the lowest SEP, outperforming both EUC-D and GAP-D by up to one order of magnitude. It can be observed that at high SNR values, the GAP-D exhibits an increase in SEP. This behavior arises because GAP-D, being insensitive to ADs, compensates phase errors by effectively increasing the number of energy levels to separate symbols with close phases. However, this results in smaller distances between the amplitude rings, rendering the constellation more susceptible to amplitude fluctuations and increasing the overall SEP. In contrast, the PAD-D adaptively balances amplitude and phase decision residuals, maintaining robustness across diverse impairment conditions and preventing the high-SNR degradation observed in GAP-D.

Fig. 4 compares the performance of the EUC-D, GAP-D, and the proposed PAD-D for SAPSK constellations with $M=\{256, 512\}$. For each SNR value, the SAPSK constellations are individually optimized for each detector to achieve the minimum attainable SEP under the corresponding detection rule. As shown in Figs. \ref{fig::detectors_comparison_sapsk}a and \ref{fig::detectors_comparison_sapsk}b, representing dominant PN and dominant AD scenarios, respectively, the proposed PAD-D consistently outperforms the other two detectors. Across all operating regimes, the PAD-D achieves a substantial reduction in SEP, often exceeding one order of magnitude, demonstrating its ability to adapt effectively to joint amplitude and phase distortions. 

\begin{figure}
	\centering
    \begin{subfigure}{\linewidth}
    \centering
    \begin{tikzpicture}
	\begin{semilogyaxis}[
    width=0.95\linewidth,
	xlabel = $\frac{E_{s}}{\sigma_n^2}$ (dB),
	ylabel = Symbol Error Probability (SEP),
	xmin = 10,
	xmax = 80,
	ymin = 0.000001,
	ymax = 1,
    xtick = {10, 20,...,80},
	grid = major,
	legend cell align = {left},
    legend pos = south west,
    legend style={font=\footnotesize}
	]

	\addplot[
	no marks,
    color=black,
    line width = 1pt,
	style = dashdotted,
	]
	table {figures/simulation_results/sapsk_512_opt_eucd_sigma_g_1e-4_sigma_phi_1e-3.txt};
    \addlegendentry{EUC-D};
    
    \addplot[
	no marks,
    color=black,
    line width = 1pt,
	style = dashed,
	]
	table {figures/simulation_results/sapsk_512_opt_gapd_sigma_g_1e-4_sigma_phi_1e-3.txt};
    \addlegendentry{GAP-D};
    
    \addplot[
	no marks,
    color=black,
    line width = 1pt,
	style = solid,
	]
	table {figures/simulation_results/sapsk_512_opt_proposed_sigma_g_1e-4_sigma_phi_1e-3.txt};
    \addlegendentry{PAD-D};

     \addplot[
	only marks,
    mark=square,
    mark options = solid,
    mark repeat = 1,
    mark size = 2.8,
    color=red,
    line width = 1pt,
	style = dashdotted,
	]
	table {figures/simulation_results/sapsk_256_opt_eucd_sigma_g_1e-4_sigma_phi_1e-3.txt};
    \addlegendentry{256-SAPSK};
    
    \addplot[
	only marks,
    mark=o,
    mark options = solid,
    mark repeat = 1,
    mark size = 2.8,
    color=black,
    line width = 1pt,
	style = dashdotted,
	]
	table {figures/simulation_results/sapsk_512_opt_eucd_sigma_g_1e-4_sigma_phi_1e-3.txt};
    \addlegendentry{512-SAPSK};

    \addplot[
	only marks,
    mark=o,
    mark options = solid,
    mark repeat = 1,
    mark size = 2.8,
    color=black,
    line width = 1pt,
	style = dashdotted,
	]
	table {figures/simulation_results/sapsk_512_opt_gapd_sigma_g_1e-4_sigma_phi_1e-3.txt};

    \addplot[
	only marks,
    mark=o,
    mark options = solid,
    mark repeat = 1,
    mark size = 2.8,
    color=black,
    line width = 1pt,
	style = dashdotted,
	]
	table {figures/simulation_results/sapsk_512_opt_proposed_sigma_g_1e-4_sigma_phi_1e-3.txt};
    
    \addplot[
	only marks,
    mark=square,
    mark options = solid,
    mark repeat = 1,
    mark size = 2.8,
    color=red,
    line width = 1pt,
	style = dashed,
	]
	table {figures/simulation_results/sapsk_256_opt_gapd_sigma_g_1e-4_sigma_phi_1e-3.txt};

    \addplot[
	only marks,
    mark=square,
    mark options = solid,
    mark repeat = 1,
    mark size = 2.8,
    color=red,
    line width = 1pt,
	style = solid,
	]
	table {figures/simulation_results/sapsk_256_opt_proposed_sigma_g_1e-4_sigma_phi_1e-3.txt};

    \addplot[
	no marks,
    color=red,
    line width = 1pt,
	style = dashdotted,
	]
	table {figures/simulation_results/sapsk_256_opt_eucd_sigma_g_1e-4_sigma_phi_1e-3.txt};

    \addplot[
	no marks,
    color=red,
    line width = 1pt,
	style = dashed,
	]
	table {figures/simulation_results/sapsk_256_opt_gapd_sigma_g_1e-4_sigma_phi_1e-3.txt};

    \addplot[
	no marks,
    color=red,
    line width = 1pt,
	style = solid,
	]
	table {figures/simulation_results/sapsk_256_opt_proposed_sigma_g_1e-4_sigma_phi_1e-3.txt};

	\end{semilogyaxis}
	\end{tikzpicture}
	\subcaption{$\sigma_g^2=10^{-4}$ and $\pn^2=10^{-3}$}
    \end{subfigure}

    \begin{subfigure}{\linewidth}
    \centering
    \begin{tikzpicture}
	\begin{semilogyaxis}[
    width=0.95\linewidth,
	xlabel = $\frac{E_{s}}{\sigma_n^2}$ (dB),
	ylabel = Symbol Error Probability (SEP),
	xmin = 10,
	xmax = 80,
	ymin = 0.000001,
	ymax = 1,
    xtick = {10, 20,...,80},
	grid = major,
	legend cell align = {left},
    legend pos = south west,
    legend style={font=\footnotesize}
	]

   	\addplot[
	no marks,
    color=black,
    line width = 1pt,
	style = dashdotted,
	]
	table {figures/simulation_results/sapsk_512_opt_eucd_sigma_g_1e-3_sigma_phi_1e-4.txt};
    \addlegendentry{EUC-D};
    
    \addplot[
	no marks,
    color=black,
    line width = 1pt,
	style = dashed,
	]
	table {figures/simulation_results/sapsk_512_opt_gapd_sigma_g_1e-3_sigma_phi_1e-4.txt};
    \addlegendentry{GAP-D};
    
    \addplot[
	no marks,
    color=black,
    line width = 1pt,
	style = solid,
	]
	table {figures/simulation_results/sapsk_512_opt_proposed_sigma_g_1e-3_sigma_phi_1e-4.txt};
    \addlegendentry{PAD-D};

     \addplot[
	only marks,
    mark=square,
    mark options = solid,
    mark repeat = 1,
    mark size = 2.8,
    color=red,
    line width = 1pt,
	style = dashdotted,
	]
	table {figures/simulation_results/sapsk_256_opt_eucd_sigma_g_1e-3_sigma_phi_1e-4.txt};
    \addlegendentry{256-SAPSK};
    
    \addplot[
	only marks,
    mark=o,
    mark options = solid,
    mark repeat = 1,
    mark size = 2.8,
    color=black,
    line width = 1pt,
	style = dashdotted,
	]
	table {figures/simulation_results/sapsk_512_opt_eucd_sigma_g_1e-3_sigma_phi_1e-4.txt};
    \addlegendentry{512-SAPSK};

    \addplot[
	only marks,
    mark=o,
    mark options = solid,
    mark repeat = 1,
    mark size = 2.8,
    color=black,
    line width = 1pt,
	style = dashdotted,
	]
	table {figures/simulation_results/sapsk_512_opt_gapd_sigma_g_1e-3_sigma_phi_1e-4.txt};

    \addplot[
	only marks,
    mark=o,
    mark options = solid,
    mark repeat = 1,
    mark size = 2.8,
    color=black,
    line width = 1pt,
	style = dashdotted,
	]
	table {figures/simulation_results/sapsk_512_opt_proposed_sigma_g_1e-3_sigma_phi_1e-4.txt};
    
    \addplot[
	only marks,
    mark=square,
    mark options = solid,
    mark repeat = 1,
    mark size = 2.8,
    color=red,
    line width = 1pt,
	style = dashed,
	]
	table {figures/simulation_results/sapsk_256_opt_gapd_sigma_g_1e-3_sigma_phi_1e-4.txt};

    \addplot[
	only marks,
    mark=square,
    mark options = solid,
    mark repeat = 1,
    mark size = 2.8,
    color=red,
    line width = 1pt,
	style = solid,
	]
	table {figures/simulation_results/sapsk_256_opt_proposed_sigma_g_1e-3_sigma_phi_1e-4.txt};

    \addplot[
	no marks,
    color=red,
    line width = 1pt,
	style = dashdotted,
	]
	table {figures/simulation_results/sapsk_256_opt_eucd_sigma_g_1e-3_sigma_phi_1e-4.txt};

    \addplot[
	no marks,
    color=red,
    line width = 1pt,
	style = dashed,
	]
	table {figures/simulation_results/sapsk_256_opt_gapd_sigma_g_1e-3_sigma_phi_1e-4.txt};

    \addplot[
	no marks,
    color=red,
    line width = 1pt,
	style = solid,
	]
	table {figures/simulation_results/sapsk_256_opt_proposed_sigma_g_1e-3_sigma_phi_1e-4.txt};
    
	\end{semilogyaxis}
	\end{tikzpicture}
	\subcaption{$\sigma_g^2=10^{-3}$ and $\pn^2=10^{-4}$}
    \end{subfigure}
    \caption{Comparison between EUC-D, GAP-D, and PAD-D for SAPSK.}
	\label{fig::detectors_comparison_sapsk}
\end{figure}

\begin{figure}
	\centering
    \begin{subfigure}{\linewidth}
    \centering
    \begin{tikzpicture}
	\begin{semilogyaxis}[
    width=0.95\linewidth,
	xlabel = $\frac{E_{s}}{\sigma_n^2}$ (dB),
	ylabel = Symbol Error Probability (SEP),
	xmin = 10,
	xmax = 80,
	ymin = 0.000001,
	ymax = 1,
    xtick = {10, 20,...,80},
	grid = major,
	legend cell align = {left},
    legend pos = south west,
    legend style={font=\footnotesize},
    legend style={at={(0,0)},anchor=south west}
	]

	\addplot[
	no marks,
    color=black,
    line width = 1.5pt,
	style = solid,
	]
	table {figures/simulation_results/SEP_approx_qam_128_sigma_g_1e-4_sigma_phi_1e-3.txt};
    \addlegendentry{SEP Approx. \eqref{eq:SEP_approx}};

    \addplot[
	only marks,
    mark=o,
    mark options = solid,
    mark repeat = 1,
    mark size = 2.8,
    color=black,
    line width = 1.5pt,
	style = dashdotted,
	]
	table {figures/simulation_results/qam_128_proposed_sigma_g_1e-4_sigma_phi_1e-3.txt};
    \addlegendentry{128-QAM};

    \addplot[
	only marks,
    mark=triangle,
    mark options = solid,
    mark repeat = 1,
    mark size = 2.8,
    color=green,
    line width = 1.5pt,
	style = dashdotted,
	]
	table {figures/simulation_results/qam_256_proposed_sigma_g_1e-4_sigma_phi_1e-3.txt};
    \addlegendentry{256-QAM};
    
    \addplot[
	only marks,
    mark=square,
    mark options = solid,
    mark repeat = 1,
    mark size = 2.8,
    color=red,
    line width = 1.5pt,
	style = dashdotted,
	]
	table {figures/simulation_results/sapsk_256_opt_proposed_sigma_g_1e-4_sigma_phi_1e-3.txt};
    \addlegendentry{256-SAPSK};

    \addplot[
	only marks,
    mark=asterisk,
    mark options = solid,
    mark repeat = 1,
    mark size = 2.8,
    color=blue,
    line width = 1.5pt,
	style = dashdotted,
	]
	table {figures/simulation_results/sapsk_512_opt_proposed_sigma_g_1e-4_sigma_phi_1e-3.txt};
    \addlegendentry{512-SAPSK};

    \addplot[
	only marks,
    mark=star,
    mark options = solid,
    mark repeat = 1,
    mark size = 2.8,
    color=brown,
    line width = 1.5pt,
	style = dashdotted,
	]
	table {figures/simulation_results/sapsk_1024_opt_proposed_sigma_g_1e-4_sigma_phi_1e-3.txt};
    \addlegendentry{1024-SAPSK};

    \addplot[
	only marks,
    mark=o,
    mark options = solid,
    mark repeat = 1,
    mark size = 2.8,
    color=black,
    line width = 1.5pt,
	style = dashdotted,
	]
	table {figures/simulation_results/qam_128_proposed_sigma_g_1e-4_sigma_phi_1e-3.txt};

    \addplot[
	no marks,
    color=green,
    line width = 1.5pt,
	style = solid,
	]
	table {figures/simulation_results/SEP_approx_qam_256_sigma_g_1e-4_sigma_phi_1e-3.txt};

    \addplot[
	no marks,
    color=red,
    line width = 1.5pt,
	style = solid,
	]
	table {figures/simulation_results/SEP_approx_sapsk_256_opt_sigma_g_1e-4_sigma_phi_1e-3.txt};

    \addplot[
	no marks,
    color=blue,
    line width = 1.5pt,
	style = solid,
	]
	table {figures/simulation_results/SEP_approx_sapsk_512_opt_sigma_g_1e-4_sigma_phi_1e-3.txt};

    \addplot[
	no marks,
    color=brown,
    line width = 1.5pt,
	style = solid,
	]
	table {figures/simulation_results/SEP_approx_sapsk_1024_opt_sigma_g_1e-4_sigma_phi_1e-3.txt};
    
	\end{semilogyaxis}
	\end{tikzpicture}
    \subcaption{$\sigma_g^2=10^{-4}$ and $\pn^2=10^{-3}$}
    \end{subfigure}
    
    \begin{subfigure}{\linewidth}
    \centering
    \begin{tikzpicture}
	\begin{semilogyaxis}[
    width=0.95\linewidth,
	xlabel = $\frac{E_{s}}{\sigma_n^2}$ (dB),
	ylabel = Symbol Error Probability (SEP),
	xmin = 10,
	xmax = 80,
	ymin = 0.000001,
	ymax = 1,
    xtick = {10, 20,...,80},
	grid = major,
	legend cell align = {left},
    legend pos = south west,
    legend style={font=\footnotesize},
    legend style={at={(0,0)},anchor=south west}
	]

	\addplot[
	no marks,
    color=black,
    line width = 1.5pt,
	style = solid,
	]
	table {figures/simulation_results/SEP_approx_qam_128_sigma_g_1e-3_sigma_phi_1e-4.txt};
    \addlegendentry{SEP Approx. \eqref{eq:SEP_approx}};

    \addplot[
	only marks,
    mark=o,
    mark options = solid,
    mark repeat = 1,
    mark size = 2.8,
    color=black,
    line width = 1.5pt,
	style = dashdotted,
	]
	table {figures/simulation_results/qam_128_proposed_sigma_g_1e-3_sigma_phi_1e-4.txt};
    \addlegendentry{128-QAM};

    \addplot[
	only marks,
    mark=triangle,
    mark options = solid,
    mark repeat = 1,
    mark size = 2.8,
    color=green,
    line width = 1.5pt,
	style = dashdotted,
	]
	table {figures/simulation_results/qam_256_proposed_sigma_g_1e-3_sigma_phi_1e-4.txt};
    \addlegendentry{256-QAM};
    
    \addplot[
	only marks,
    mark=square,
    mark options = solid,
    mark repeat = 1,
    mark size = 2.8,
    color=red,
    line width = 1.5pt,
	style = dashdotted,
	]
	table {figures/simulation_results/sapsk_256_opt_proposed_sigma_g_1e-3_sigma_phi_1e-4.txt};
    \addlegendentry{256-SAPSK};

    \addplot[
	only marks,
    mark=asterisk,
    mark options = solid,
    mark repeat = 1,
    mark size = 2.8,
    color=blue,
    line width = 1.5pt,
	style = dashdotted,
	]
	table {figures/simulation_results/sapsk_512_opt_proposed_sigma_g_1e-3_sigma_phi_1e-4.txt};
    \addlegendentry{512-SAPSK};

    \addplot[
	only marks,
    mark=star,
    mark options = solid,
    mark repeat = 1,
    mark size = 2.8,
    color=brown,
    line width = 1.5pt,
	style = dashdotted,
	]
	table {figures/simulation_results/sapsk_1024_opt_proposed_sigma_g_1e-3_sigma_phi_1e-4.txt};
    \addlegendentry{1024-SAPSK};

    \addplot[
	only marks,
    mark=o,
    mark options = solid,
    mark repeat = 1,
    mark size = 2.8,
    color=black,
    line width = 1.5pt,
	style = dashdotted,
	]
	table {figures/simulation_results/qam_128_proposed_sigma_g_1e-3_sigma_phi_1e-4.txt};

    \addplot[
	no marks,
    color=green,
    line width = 1.5pt,
	style = solid,
	]
	table {figures/simulation_results/SEP_approx_qam_256_sigma_g_1e-3_sigma_phi_1e-4.txt};

    \addplot[
	no marks,
    color=red,
    line width = 1.5pt,
	style = solid,
	]
	table {figures/simulation_results/SEP_approx_sapsk_256_opt_sigma_g_1e-3_sigma_phi_1e-4.txt};

    \addplot[
	no marks,
    color=blue,
    line width = 1.5pt,
	style = solid,
	]
	table {figures/simulation_results/SEP_approx_sapsk_512_opt_sigma_g_1e-3_sigma_phi_1e-4.txt};

    \addplot[
	no marks,
    color=brown,
    line width = 1.5pt,
	style = solid,
	]
	table {figures/simulation_results/SEP_approx_sapsk_1024_opt_sigma_g_1e-3_sigma_phi_1e-4.txt};
	\end{semilogyaxis}
	\end{tikzpicture}
    \subcaption{$\sigma_g^2=10^{-3}$ and $\pn^2=10^{-4}$}
    \end{subfigure}
    \caption{SEP approximation for PAD-D across various constellations.}
	\label{fig::SEP_approx}
\end{figure}
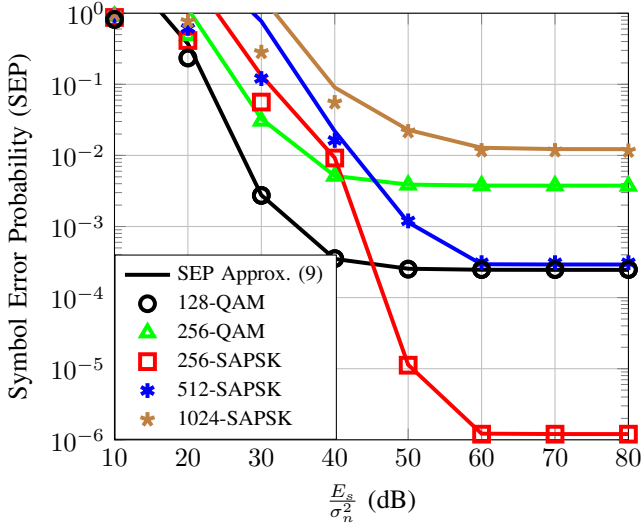
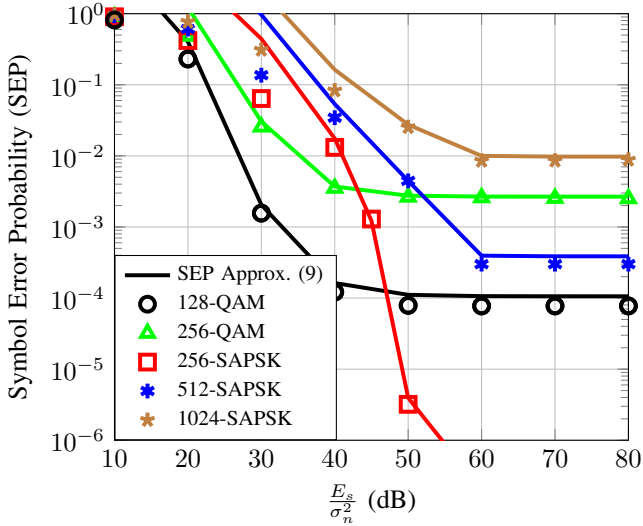

Fig. \ref{fig::SEP_approx} validates the accuracy and generality of the proposed closed-form SEP approximation for the PAD-D under two distinct HW impairment conditions, shown in Figs. \ref{fig::SEP_approx}a and \ref{fig::SEP_approx}b. The analytical expression shows excellent agreement with the simulated results across both standard high-order QAM and SNR-optimized SAPSK constellations, confirming its applicability to various modulation formats. In both scenarios, the approximation remains tight in the high-SNR region, where the Gaussian moment-matching assumptions hold, and serves as a consistent upper bound at lower SNRs. These results demonstrate that the proposed closed-form formulation offers an accurate and analytically convenient tool for performance analysis and system-level optimization across various HW impairment conditions.

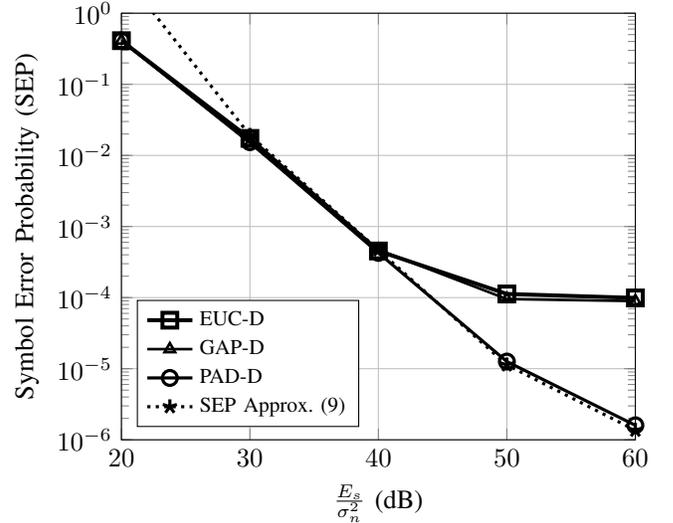
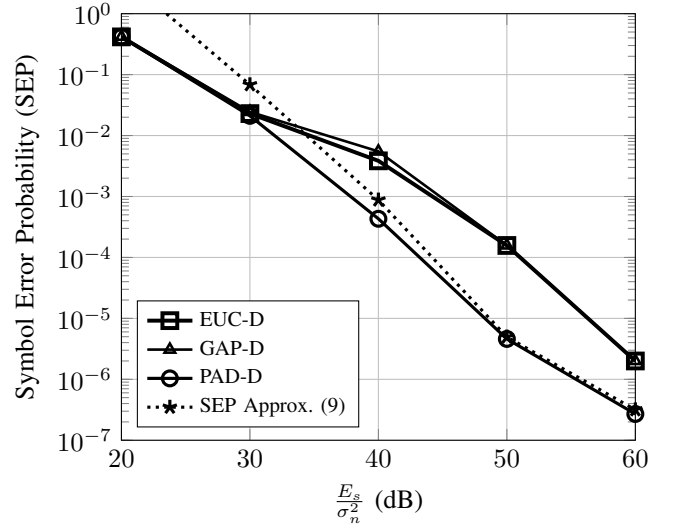
\begin{figure}
	\centering
    \begin{subfigure}{\linewidth}
    \centering
    \begin{tikzpicture}
	\begin{semilogyaxis}[
    width=0.95\linewidth,
	xlabel = $\frac{E_{s}}{\sigma_n^2}$ (dB),
	ylabel = Symbol Error Probability (SEP),
	xmin = 20,
	xmax = 60,
	ymin = 0.000001,
	ymax = 1,
    xtick = {20,30,...,60},
	grid = major,
	legend cell align = {left},
    legend pos = south west,
    legend style={font=\footnotesize}
	]

    \addplot[
    mark=square,
    mark options = solid,
    mark repeat = 1,
    mark size = 2.8,
    color=black,
    line width = 1.5pt,
	style = solid,
	]
	table {figures/simulation_results/optimized/opt_256_eucd_sigma_g_1e-4_sigma_phi_1e-3.txt};
    \addlegendentry{EUC-D};
    
    \addplot[
    mark=triangle,
    mark options = solid,
    mark repeat = 1,
    mark size = 2.3,
    color=black,
    line width = 1pt,
	style = solid,
	]
	table {figures/simulation_results/optimized/opt_256_gapd_sigma_g_1e-4_sigma_phi_1e-3.txt};
    \addlegendentry{GAP-D};

    \addplot[
    mark=o,
    mark options = solid,
    mark repeat = 1,
    mark size = 2.8,
    color=black,
    line width = 1.2pt,
	style = solid,
	]
	table {figures/simulation_results/optimized/opt_256_proposed_sigma_g_1e-4_sigma_phi_1e-3.txt};
    \addlegendentry{PAD-D};

    \addplot[
      mark=star,
      mark options={solid, fill=black},
      mark size=2.8,
      color=black,
      line width=1.2pt,
      style=dotted,
      ]
	table {figures/simulation_results/optimized/SEP_approx_SN_opt_256_sigma_g_1e-4_sigma_phi_1e-3.txt};
    \addlegendentry{SEP Approx. \eqref{eq:SEP_approx}};

	\end{semilogyaxis}
	\end{tikzpicture}
    \subcaption{$\sigma_g^2=10^{-4}$ and $\pn^2=10^{-3}$}
    \end{subfigure}
    
    \begin{subfigure}{\linewidth}
    \centering
    \begin{tikzpicture}
	\begin{semilogyaxis}[
    width=0.95\linewidth,
	xlabel = $\frac{E_{s}}{\sigma_n^2}$ (dB),
	ylabel = Symbol Error Probability (SEP),
	xmin = 20,
	xmax = 60,
	ymin = 0.0000001,
	ymax = 1,
    xtick = {20, 30,...,60},
	grid = major,
	legend cell align = {left},
    legend pos = south west,
    legend style={font=\footnotesize}
	]
    \addplot[
    mark=square,
    mark options = solid,
    mark repeat = 1,
    mark size = 2.8,
    color=black,
    line width = 1.5pt,
	style = solid,
	]
	table {figures/simulation_results/optimized/opt_256_eucd_sigma_g_1e-3_sigma_phi_1e-4.txt};
    \addlegendentry{EUC-D};
    
    \addplot[
    mark=triangle,
    mark options = solid,
    mark repeat = 1,
    mark size = 2.3,
    color=black,
    line width = 1pt,
	style = solid,
	]
	table {figures/simulation_results/optimized/opt_256_gapd_sigma_g_1e-3_sigma_phi_1e-4.txt};
    \addlegendentry{GAP-D};

    \addplot[
    mark=o,
    mark options = solid,
    mark repeat = 1,
    mark size = 2.8,
    color=black,
    line width = 1.2pt,
	style = solid,
	]
	table {figures/simulation_results/optimized/opt_256_proposed_sigma_g_1e-3_sigma_phi_1e-4.txt};
    \addlegendentry{PAD-D};

    \addplot[
      mark=star,
      mark options={solid, fill=black},
      mark size=2.8,
      color=black,
      line width=1.2pt,
      style=dotted,
      ]
	table {figures/simulation_results/optimized/SEP_approx_SN_opt_256_sigma_g_1e-3_sigma_phi_1e-4.txt};
    \addlegendentry{SEP Approx. \eqref{eq:SEP_approx}};
	\end{semilogyaxis}
	\end{tikzpicture}
    \subcaption{$\sigma_g^2=10^{-3}$ and $\pn^2=10^{-4}$}
    \end{subfigure}
    \caption{Constellation optimization for $M=256$.}
	\label{fig::optimized_constelations_with_SEP_approx}
\end{figure}

Finally, Fig. \ref{fig::optimized_constelations_with_SEP_approx} illustrates the optimized constellations and corresponding SEP performance obtained by independently solving the optimization problem $(\textbf{P1})$ for each SNR value under the EUC-D, GAP-D, and PAD-D detection metrics. The optimized constellations for each detector are then evaluated via Monte Carlo simulations to estimate their achievable SEPs. As shown, the proposed PAD-D consistently yields the lowest SEP across all operating conditions, clearly outperforming both EUC-D and GAP-D in scenarios affected by AD and PN. In contrast, the EUC-D and GAP-D approaches exhibit limited robustness, with the former neglecting transceiver distortions entirely and the latter accounting only for PN while remaining insensitive to AD. Furthermore, the close agreement between the simulated SEPs and the analytical approximations confirms the validity and tightness of the derived high-SNR expression across all optimized constellations. This demonstrates that the analytical SEP model remains accurate and applicable even for geometries obtained through constellation optimization.
\vspace{-1.6em}
\section{Conclusion}
This paper introduces a novel maximum-likelihood detection framework that considers HW impairments in both amplitude and phase. The proposed PAD-D operates in the polar domain and mitigates amplitude and phase decision residuals through distortion-aware weighting. Unlike conventional detectors, the PAD-D performs robustly under generalized transceiver impairment conditions, outperforming the EUC-D and GAP-D. A tight closed-form high-SNR SEP approximation for high-order constellations was also derived, providing a generic analytical tool applicable to arbitrary constellations. Building on this framework, optimized constellations were further developed to minimize the average SEP under the PAD-D metric, where symbol positions are globally refined through simulated annealing combined with gradient-based optimization. Simulation results confirmed that the PAD-D significantly reduces the error floor by up to an order of magnitude compared to the EUC-D and GAP-D for both high-order QAM and optimized constellations, while the optimized constellations themselves provide additional SEP gains over conventional designs by better adapting the symbol geometry to the underlying distortion characteristics, and the analytical SEP approximation remains accurate even for these optimized configurations. Beyond its empirical performance, the analytical and optimization framework provides valuable insights for system-level adaptation, adaptive constellation shaping, and the design of next-generation transceivers resilient to combined amplitude and phase HW impairments.
\appendices
\section{Proof of Theorem \ref{prop:detection_metric}}
The received symbol is given by \eqref{eq:system_model}. Rotate the observation by $-(\phi + \arg\{s_m\})$ and define $n' = ne^{-j(\phi+\arg\{s\})}$ which follows $\mathcal{CN}(0,\sigma_n^2)$. 
The AD can be written as $g=1+\epsilon_g$ with $\epsilon_g\sim \mathcal{N}(0, \sigma_g^2)$ and by rotating the received symbol, we obtain 
\begin{equation}
    \begin{aligned}
        re^{-j(\phi + \arg\{s_m\}))} = (1+\epsilon_g)|s_m| + n'.
    \end{aligned}
\end{equation}
Thus, the magnitude of the received signal is given by 
\begin{equation}
    \begin{aligned}
        |r| &= \left|(1+\epsilon_g)|s_m| + n'\right|\\
        &= \sqrt{\left((1+\epsilon_g)|s_m| + \Re\{n'\}\right)^2 + (\Im\{n'\})^2}\\
        &\approx (1+\epsilon_g)|s_m| + \Re\{n'\},
    \end{aligned}
\end{equation}
where $\Re(\cdot)$ and $\Im(\cdot)$ denote the real and imaginary parts of a complex number, respectively. In the last step, high-SNR is assumed, and thus the amplitude decision residual is given by
\begin{equation}\label{eq:amplitude_residual}
    \begin{aligned}
        |r| - |s_m| \approx\epsilon_g|s_m| + \Re\{n'\}.
    \end{aligned}
\end{equation}
Similarly, in the high-SNR regime, the phase decision residual follows the familiar small-angle expansion, and it is written as in \cite{soft_metrics}
\begin{equation}\label{eq:phase_residual}
    \begin{aligned}
        \arg\{r\}-\arg\{s_m\} \approx \phi+ \frac{\Im\{n'\}}{|s_m|}.
    \end{aligned}
\end{equation}
Since $\epsilon_g$, $\phi$, and the real and imaginary parts of $n'$ are zero-mean Gaussian independent random variables, the amplitude and phase decision residuals in \eqref{eq:amplitude_residual} and \eqref{eq:phase_residual}, respectively, are Gaussian random variables with zero mean and variances 
\begin{equation}
    \begin{aligned}
        V_{m}^{(a)} = \frac{\sigma_n^2}{2} + \sigma_g^2|s_m|^2
    \end{aligned}
\end{equation}
and
\begin{equation}
    \begin{aligned}
        V_{m}^{(\theta)} = \pn^2 + \frac{\sigma_n^2}{2|s_m|^2},
    \end{aligned}
\end{equation}
respectively.
Therefore, the conditional density $p(|r|,\arg\{r\}|s)$ is given by 
\begin{equation}
    \begin{aligned}
    p(|r|,\arg\{r\}|s_m) =
    \frac{\exp\left(-\frac{(|r|-|s_m|)^2}{2V_{m}^{(a)}} \!-\! \frac{(\arg\{r\}-\arg\{s_m\})^2}{2V_\theta(s)}\right)}{2\pi\sqrt{V_{m}^{(a)}V_{m}^{(\theta)}}},
    \end{aligned}
\end{equation}
and the negative log-likelihood yields the PAD-D metric in \eqref{eq:L}
which concludes the proof.
\vspace{-1em}
\section{Proof of Theorem \ref{prop:SEP_approx}}
We consider the pairwise error probability between two candidate symbols $s_i$ and $s_j$ under the PAD-D detection metric.  
Let $\eta_{ij}$ denote the log-likelihood difference, i.e.,
\begin{equation}
    \eta_{ij} \triangleq L_j - L_i,
\end{equation}
which, conditioned on $s_i$ being transmitted, can be expressed as
\begin{equation}
    \eta_{ij} = a_0 w^2 + 2 a_1 w + a_2\psi^2 +2 a_3 \psi + a_4,
\end{equation}
where $w \sim \mathcal{N}(0,V_i^{(a)})$ and $\psi \sim \mathcal{N}(0,V_i^{(\theta)})$.  
The coefficients $a_0$, $a_1$, $a_2$, $a_3$, and $a_4$ are functions of the constellation radii and phases, as well as the parameters $V_i^{(a)}$, $V_j^{(a)}$, $V_i^{(\theta)}$, and $V_j^{(\theta)}$, and they are written as 
\begin{equation*}
    \begin{aligned}
        a_0 &= \frac{1}{V_j^{(a)}} - \frac{1}{V_i^{(a)}}, \hspace{2em}
    a_1 = \frac{|s_j|-|s_i|}{V_j^{(a)}}, \\
    a_2 &= \frac{1}{V_j^{(\theta)}} - \frac{1}{V_i^{(\theta)}}, \hspace{2em}
    a_3 = \frac{\arg(s_j)-\arg(s_i)}{V_j^{(\theta)}}, \\
    a_4 &= \frac{(|s_j|-|s_i|)^2}{V_j^{(a)}} 
       \!+\! \frac{(\arg(s_j)-\arg(s_i))^2}{V_j^{(\theta)}} \!+\! \ln\left(\frac{V_j^{(a)} V_j^{(\theta)}}{V_i^{(a)} V_i^{(\theta)}}\right)\! .
    \end{aligned}
\end{equation*}
In the high-SNR regime, $a_2\rightarrow0$ because $V_j^{(\theta)} \approx  V_i^{(\theta)} \approx \pn^2$, and thus $\eta_{ij}$ can be rewritten as
\begin{equation}\label{eq:eta_ij}
    \eta_{ij} = a_0 w^2 + 2 a_2 w + 2 a_3 \psi + a_4.
\end{equation}
Furthermore, considering that $\mathbb{E}[w] = 0$, $\mathbb{E}[w^2] = V_i^{(a)}$, 
$\mathrm{Var}(w^2) = 2\left(V_i^{(a)}\right)^2$, $\mathbb{E}[\psi]=0$, and $\mathrm{Var}(\psi)=V_i^{(\theta)}$, the first three moments of $\eta_{ij}$ are given in closed form as
\begin{equation}
    \begin{aligned}
        \mu_{ij} = a_0 V_i^{(a)} + a_4,
    \end{aligned}
\end{equation}
\begin{equation}
     \begin{aligned}
         \sigma_{ij}^2 = 2 a_0^2 \left(V_i^{(a)}\right)^2 + 4 a_1^2 V_i^{(a)} + 4 a_3^2 V_i^{(\theta)},
     \end{aligned}
 \end{equation}
and 
\begin{equation}
     \begin{aligned}
        \gamma_{1,ij} = \frac{8a_0^3\left(V_i^{(a)}\right)^3 + 24a_0a_1^2\left(V_i^{(a)}\right)^2}{\sigma_{ij}^3},
     \end{aligned}
 \end{equation}
where $\gamma_{1,ij}$ is the standardized skewness.

Moreover, it follows by \eqref{eq:eta_ij} that the source of non-Gaussianity is governed by the quadratic coefficient $a_0$, which is only a function of the parameters $V_i^{(a)}$ and $V_j^{(a)}$.
For neighboring symbols with equal amplitude, these parameters are identical, yielding $a_0=0$, thus $\eta_{ij}$ reduces exactly to a linear Gaussian form. For the rest neighbors, as the constellation order $M$ increases and the amplitude rings densify, the dominant nearest-neighbor pairs exhibit a shrinking gap $|V_j^{(a)} - V_i^{(a)}|$, which drives $a_0$ to be small and keeps the induced asymmetry of $\eta_{ij}$ small, motivating an approximation that retains mean and variance while also reflecting a mild but non-negligible skewness using a single extra shape parameter.

Therefore, we approximate $\eta_{ij}$ by a skew-normal random variable $\mathrm{SN}\left(\xi_{ij}, \omega_{ij}, \alpha_{ij}\right)$ whose parameters are determined by matching the triplet $\left(\mu_{ij}, \sigma_{ij}, \gamma_{1,ij}\right)$. For notational convenience, we use the standard reparameterization $\delta_{ij} = \alpha_{ij}/\sqrt{1+\alpha_{ij}}\in(-1,1)$, so that the mean, variance, and standardized skewness of a skew-normal variable are given by
\begin{equation}
    \begin{aligned}
        \mu_{\text{SN}} = \xi +\omega\delta\sqrt{\frac{2}{\pi}},
    \end{aligned}
\end{equation}
\begin{equation}
    \begin{aligned}
        \sigma_{\text{SN}} = \omega^2\left(1-\frac{2\delta^2}{\pi}\right),
    \end{aligned}
\end{equation}
and 
\begin{equation}
    \begin{aligned}
        \gamma_{1,\text{SN}}(\delta) = \frac{\left(4-\pi\right)\left(\delta\sqrt{2/\pi}\right)^3}{2\left(1-2\delta^2/\pi\right)^{3/2}}.
    \end{aligned}
\end{equation}
Because $\gamma_{1,\text{SN}}(\delta)$ is continuous and strictly increasing on $(-1,1)$, there exists a unique $\delta_{ij}\in(-1,1)$ such that $\gamma_{1,\text{SN}}(\delta_{ij}) = \gamma_{1,ij}$, from which $\delta_{ij}$ can be determined with simple algebraic manipulations as in \eqref{eq:delta_ij}.
Following that, matching the mean and the variance yields 
\begin{equation}\label{eq:om}
    \begin{aligned}
        \omega_{ij} = \frac{\sigma_{ij}}{\sqrt{1-2\delta_{ij}^2/\pi}},
    \end{aligned}
\end{equation}
\begin{equation}\label{eq:x}
    \begin{aligned}
        \xi_{ij} = \mu_{ij} - \omega_{ij}\delta_{ij}\sqrt{2/\pi},
    \end{aligned}
\end{equation}
and 
\begin{equation}\label{eq:a}
    \begin{aligned}
        \alpha_{ij} = \frac{\delta_{ij}}{\sqrt{1-\delta_{ij}^2}}.
    \end{aligned}
\end{equation}
Consequently, using the CDF of the skew-normal distribution function given by
\begin{equation}
    \begin{aligned}
        F_{\mathrm{SN}}(x)
= 1 - Q\left(\frac{x-\xi}{\omega}\right)
- 2T\left(\frac{x-\xi}{\omega},\alpha\right),
    \end{aligned}
\end{equation}
as well as \eqref{eq:om}, \eqref{eq:x}, and \eqref{eq:a}, the pairwise error probability can be approximated as
\begin{equation}
    \begin{aligned}
 \Pr\{\eta_{ij}<0\}\approx Q\left(\frac{\xi_{ij}}{\omega_{ij}}\right)
 - 2T\left(\frac{-\xi_{ij}}{\omega_{ij}},\alpha_{ij}\right),
    \end{aligned}
\end{equation}
which completes the proof.
\vspace{-1em}
\bibliographystyle{IEEEtran}
\bibliography{bibliography}
\end{document}